\documentclass{article}

\usepackage[final,nonatbib]{neurips_2019}

\usepackage[utf8]{inputenc} % allow utf-8 input
\usepackage[T1]{fontenc}    % use 8-bit T1 fonts
\usepackage{hyperref}       % hyperlinks
\usepackage{url}            % simple URL typesetting
\usepackage{booktabs}       % professional-quality tables
\usepackage{amsfonts}       % blackboard math symbols
\usepackage{nicefrac}       % compact symbols for 1/2, etc.
\usepackage{microtype}      % microtypography
\usepackage{graphicx}
\usepackage{amsmath}
\usepackage{svg}
\usepackage{enumitem}
\usepackage{cleveref}

\title{Quantized deep learning models on low-power edge devices for robotic systems}

\author{%
  Anugraha Sinha \\ 
  MLT Labs, Japan \\
  \And Naveen Kumar \\
  MLT Labs, Japan \\
  \And Murukesh Mohanan \\
  MLT Labs, Japan \\
   \\ \And MD Muhaimin Rahman \\
   MLT Labs, Japan \\
 \\ \AND Yves Quemener \\
  Hackerfarm, Japan \\
 \\ \And Amina Mim \\
  MLT Labs, Japan \\
  \And Suzana Ilić\textsuperscript{1} \\
  MLT Labs, Japan \\
}

\begin{document}
\footnote{MLT Labs, Tokyo, Japan. Correspondence to: Suzana Ilić <suzana@mltokyo.ai>.}

\maketitle

\begin{abstract}
  In this work, we present a quantized deep neural network deployed on a low-power edge device, inferring learned motor-movements of a suspended robot in a defined space. This serves as the fundamental building block for the original setup, a robotic system for farms or greenhouses aimed at a wide range of agricultural tasks. Deep learning on edge devices and its implications could have a substantial impact on farming systems in the developing world, leading not only to sustainable food production and income, but also increased data privacy and autonomy.
\end{abstract}

\section{Introduction}

Agricultural technology offers the potential to support and increase farm productivity and income in the developing world \cite{feder1985adoption}. Large-scale studies have shown that adopted technologies play a substantial role in food production increases of farming systems and their impact on environmental and social outcomes \cite{pretty2006resource}. However, the adoption of new technology is constrained due to economic factors, unstable infrastructure, limited access to information and the dependency on expensive or sometimes inaccessible prerequisite hardware, software and broadband. To support the adoption of agricultural technology we propose a system that can be deployed in a wide range of agricultural contexts (e.g. 'pick and place' \cite{rodriguez2013grasping}): A neural network-driven robotic system trained on data based on a simplified mathematical reconstruction of a 3D space and ported to an edge device to enable learned rather than static motor-movements. Deep learning on edge devices has major benefits that could have great impact on the developing world:

\begin{itemize}[noitemsep,topsep=0pt,parsep=0pt,partopsep=0pt]
    \item Low latency: Fast on-device inference
    \item Privacy: Data is processed on-device
    \item Connectivity: Fully offline
    \item Power consumption: Low-power, low-cost
\end{itemize}

We utilize a device that is available at a very low cost (approximately \$15 per microcontroller) and requires low power consumption in working mode. Data processing and inference are executed on-device, which facilitates autonomy and data privacy without the need for a broadband connection, server round-trips or cloud services. In addition, a deep learning model provides the opportunity for transfer learning and continuous data integration depending on the individual farm or greenhouse, including changes in the environment and hardware constraints. Thus, these factors make quantized deep learning models on edge devices a fundamental building block for supporting agricultural sustainability in developing countries.

\section{Related work}

Deep learning in edge computing environments has recently gained traction with resource-efficiency and performance optimization of IoT devices \cite{li2018learning, chen2018thriftyedge} and wireless communication in edge learning \cite{zhu2018towards}. Previous research efforts to improve cable-suspended robots with neural networks include solving the forward kinematics \cite{inbook} and adaptive neural trajectory tracking controllers for cable-driven parallel robots \cite{Jabbari2017AdaptiveNN} validated through simulation. We combine several technologies using a neural network model to predict motor-movements of a point robot in simulation and port it for inference to an edge device with similar model performance on-device.

\section{Experimental setup}
The following sections describe the experimental setup including simulation environment and data generation, the deep learning architecture, as well as the conversion and quantization \cite{jiang2018efficient} of this model in order to be ported to the edge device, predicting the motor movements of a suspended robot in a defined perimeter. 

\subsection{Scenario description}

\begin{figure}
  \includegraphics[width=0.5\textwidth]{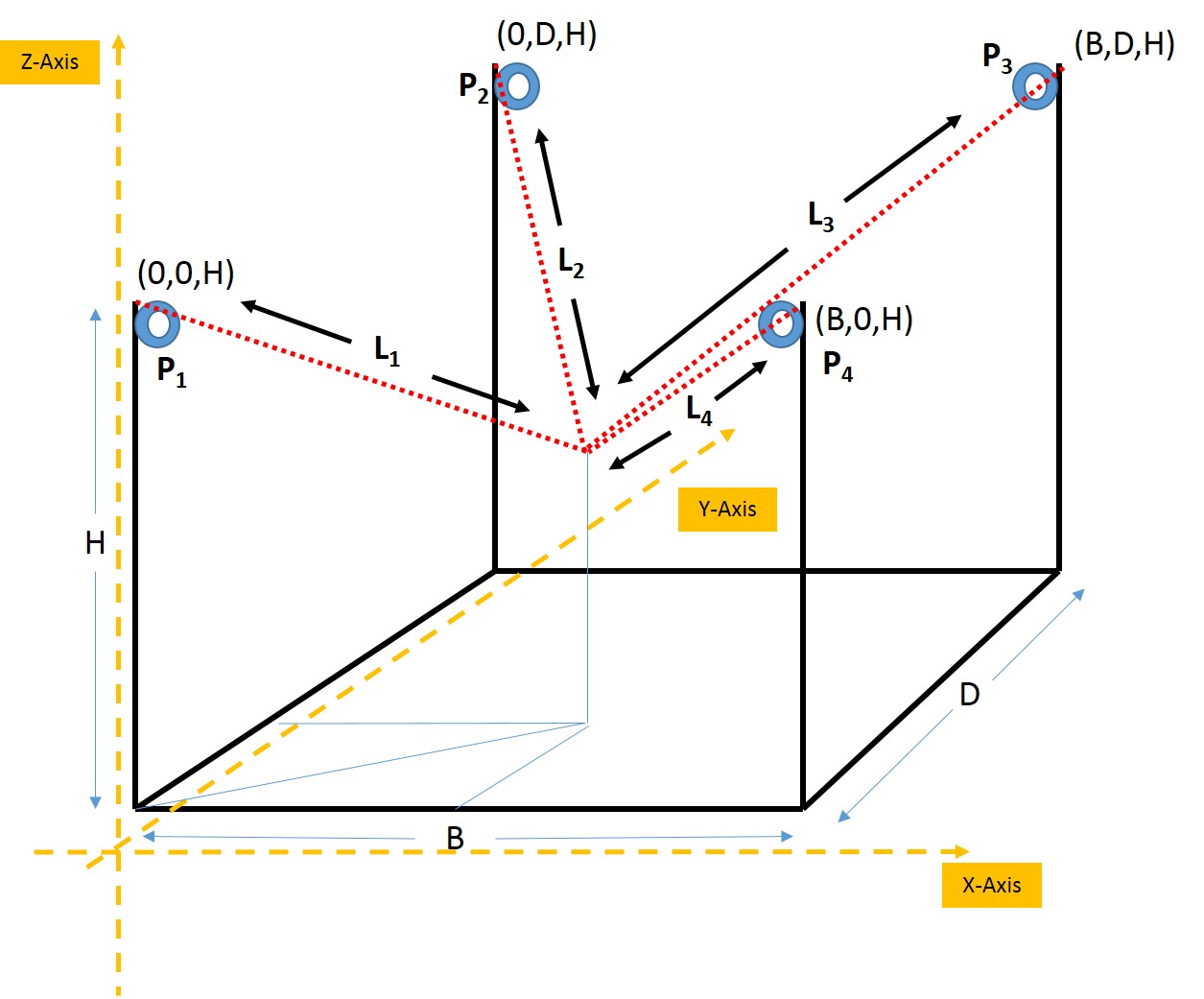}
  \centering
  \caption{Simulated environment setup.}
  \label{fig:scenario}
\end{figure}

The perimeter structure has four poles, with one pulley on top of each pole. The suspension of the robot will be through cables or strings attached to the robot wired around pulleys at each of the four corners, where each pulley is connected to a motor. A simplified reconstruction of this setup is depicted in \cref{fig:scenario}. The simulation environment is modeled after the open source system \emph{Faebot}, a cable-suspended robot for automating agricultural tasks.\footnote{\href{https://fae-bot.org/}{\url{https://fae-bot.org/}}} 
Here, the dimensions of the setup are defined as
\begin{itemize}[noitemsep,topsep=2pt,parsep=2pt,partopsep=2pt]
    \item \textit{B = Breadth of the perimeter (Distance between the poles)}
    \item \textit{H = Height of the perimeter (Height of mounted pulleys - assumed same for all)}
    \item \textit{D = Depth of the perimeter (Distance between the poles)}
    \item \textit{$L_{i}$ = Length of the string $i$, $i \in [1,4]$} 
\end{itemize}
Based on the above, we can define coordinates for the four pulleys as
\begin{align*}
    P_{1} = (0, 0, H) &&
    P_{2} = (0, D, H) &&
    P_{3} = (B, D, H) &&
    P_{4} = (B, 0, H)
\end{align*}

Using standard Cartesian geometry, we can describe different elements of the setup:
%\newpage
Relationship between current string length ($L_i$) and initial string length ($L_i^0$):
\begin{align}
L_{i}=L_{i}^0 + r_{i} \Delta \theta_{i}
\label{eqn:lengths}
\end{align}
where $r_i$ is the radius of the pulley on $P_i$, and $\Delta \theta_i$ is the rotation that it has undergone. Throughout this paper, we will be treating the robot as a point, the strings will be attached to this point. In the supplementary \cref{supp}, a further extension of the robot is depicted.

Relationship between the coordinates $(x,y,z)$ of the robot with the string lengths:
\begin{equation}\label{stringeq}
\begin{aligned}
L_1^2 &= x^2 + y^2 + (H - z)^2 \\
L_2^2 &= x^2 + (D - y)^2 + (H - z)^2 \\
L_3^2 &= (B - x)^2 + (D - y)^2 + (H - z)^2 \\
L_4^2 &= (B - x)^2 + y^2 + (H - z)^2 \\
\end{aligned}
\end{equation}

\subsection{Data generation}
Prior to deploying a system in a real-world setup, we use a simple mathematical model to generate data and test the cycle of training and porting, converting and quantizing the saved model to the microcontroller for inference. The task expects the point robot to move from a $(x_i,y_i,z_i)$ to another coordinate $(x_f,y_f,z_f)$. As the robot is suspended by four cables, all four motors need to move in synchronous manner for the robot to travel from origin to destination. We assume that the base location of the robot is at the top center of the structure, i.e., $(B/2 , D/2, H)$. All movements have the base location as the starting coordinates. Subsequently, $\theta_i$ with $i \in [0,3]$ for such movement can be described as
\begin{align}
    \Delta\theta_i = f(x_f,y_f,z_f,B,D,H,R) &&
    i \in [0,3]
\end{align}
Using \cref{stringeq} we can calculate the initial and final lengths of the strings $L_i^0$ and $L_i$. Thereafter, using \cref{eqn:lengths} we can calculate the $\Delta\theta_i$ which the motors need to move. For generating data, dimensional equality was taken as \textbf{$1 unit = 1 meter$}. Also, following ranges were considered for $(B,D,H,R)$ variables:
\begin{align*}
    B = D = H = range(1,7,0.5) &&
    R = [0.008,0.009,0.010]
\end{align*}
All possible combinations of $(B,D,H,R)$ can be represented as $(b,d,h,r)$, and to simulate movements over maximum region in space, following range was considered for $(x_f,y_f,z_f)$
\begin{align*}
    x = range(0,b,0.5) &&
    y = range(0,d,0.5) &&
    z = range(0,h,0.5) &&
    r = R
\end{align*}
Using the above data generation methodology, the dataset generated had 110,592 rows with 6 features/independent variables. The output consists of 4 values related to individual $\Delta\Theta_i$. The distribution of $\Delta\Theta_i$ can be seen in Figure \ref{fig:delta}.

\subsection{Model training and evaluation}
We implemented baseline algorithms and a mutli-layer neural network for a multi-dimensional regression task to predict four $\theta$ values at a time. These values correspond to the motor-movements in each corner of the setup. We compare the results of classical machine learning algorithms to the performance of a multi-layer neural network in Table \ref{tbl:results}. Apart from superior performance for all predicted values on the test set, deep learning models offer the opportunity for transfer learning and a feedback loop leading to a self-correcting system. These properties are essential in complex agricultural environments that suffer from extreme weather conditions or other dynamic constraints.
%Even though it's possible to port conventional machine %learning models to Micro Controller Units (MCUs), popular %libraries do not provide a framework or tools to enable %such porting. On the contrary, TensorFlow-based deep %learning models can be ported to MCUs using TensorFlow Lite micro.
%Apart from providing application-based benefits, deep learning algorithms provide more generalized models. The error distribution of deep learning models is more unified and consistent when compared with the best performing standard machine learning model (GradBoost), see appendix section \ref{gradboost}.

\subsubsection{Modelling conclusion}
\begin{itemize}[noitemsep,topsep=0pt,parsep=0pt,partopsep=0pt]
    \item Similar performance $(r2 score)$ of complex models and simpler models
    \item Generalizability of NN models is better in comparison to standard ML models (based error distribution)
    \item NN models provide ready-to-use porting frameworks to MCUs, providing better viability towards low-cost edge based predictions
\end{itemize}

\subsection{Deep learning inference at the edge}
\begin{figure}[h!]
  \centering
  \includegraphics[width=1\textwidth]{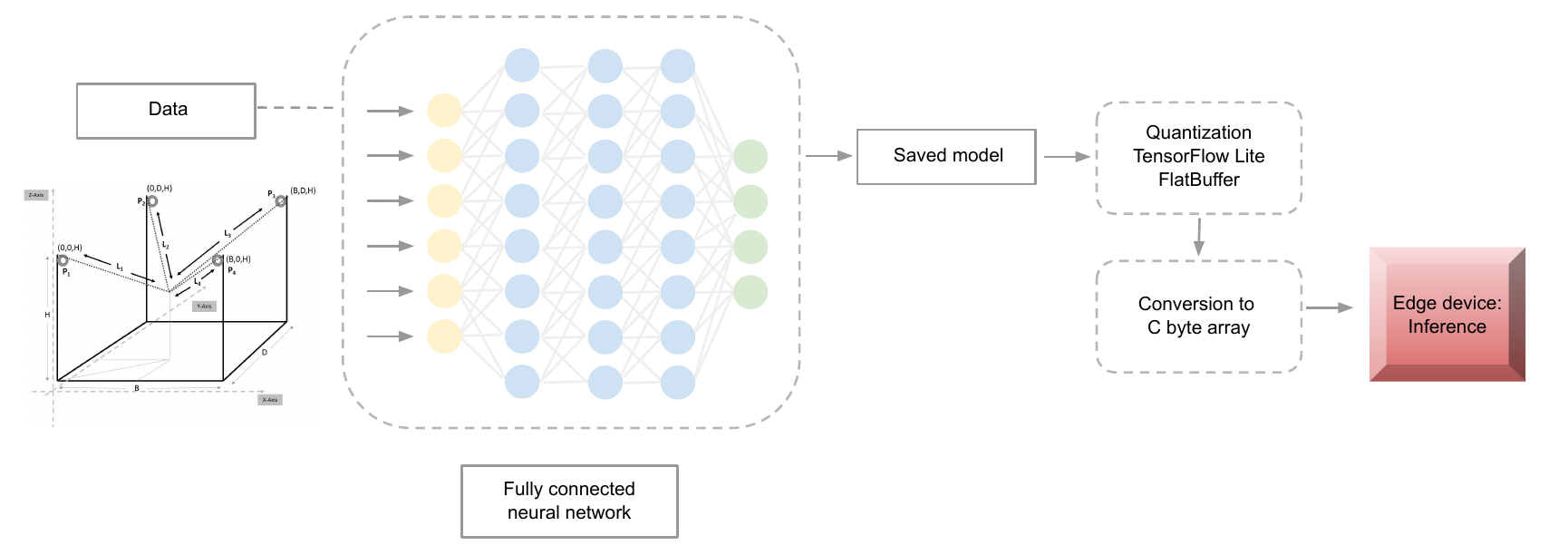}
  \caption{Data generation, model training, quantization, conversion and deployment on device.}
  \label{fig:edge}
\end{figure}

The model was trained and saved using the Keras API.\footnote{\href{https://keras.io/}{\url{https://keras.io/}}} For conversion and quantization we used TensorFlow Lite,\footnote{\href{https://www.tensorflow.org/lite}{\url{https://www.tensorflow.org/lite}}} an open source deep learning framework for on-device inference, based on TensorFlow\cite{abadi2016tensorflow}. The saved model was loaded and converted to a TensorFlow Lite FlatBuffer file using a post-training weight quantization conversion technique. This technique quantizes only the weights from floating point to 8-bits of precision. The FlatBuffer file was converted to a C byte array and saved in the form of a C source file. The converted model was then deployed to the device and run locally using the TensorFlow Lite interpreter (see \cref{fig:edge}). During inference, weights are converted from 8-bits of precision to floating point and computed using floating-point kernels. The deployed model size was 65KB. Despite the significant change in size, model performance after quantization and conversion has only slightly decreased compared to the full-size GPU-trained model. Data and code are available online.~\footnote{\href{https://github.com/Machine-Learning-Tokyo/Agritech}{\url{https://github.com/Machine-Learning-Tokyo/Agritech}}}.

\paragraph{Hardware and SDK details}

\begin{itemize}
    \item Development Board: SparkFun Edge
    \item Processor: 32-bit ARM Cortex-M4F 
    \item CPU clock: 48MHz (96MHz burst mode)
    \item Power usage: 3mA at 48MHz
    \item SRAM: 384KB 
    \item Flash memory: 1 MB
    \item Ambiq Apollo3 SDK
    \item TensorFlow Lite for Microcontrollers C++ library
\end{itemize}

\section{Conclusion and future directions}
Deep learning on edge devices can be a game-changing technology for developing countries, keeping the cost, power consumption and latency extremely low, while preserving data security and privacy. In this paper we showed how a mathematical reconstruction of a robotic system can be used to generate training data for a neural network that predicts motor-movements in a simplified environment. After quantization and conversion the model was deployed on a low-power edge device for inference without significant performance loss. Next steps involve the deployment in the real-world scenario and integrating an error-correcting feedback loop into the system.

%\begin{table}
%  \caption{Sample table title}
%  \label{sample-table}
%  \centering
%  \begin{tabular}{lll}
%    \toprule
%    \multicolumn{2}{c}{Part}                   \\
%    \cmidrule(r){1-2}
%    Name     & Description     & Size ($\mu$m) \\
%    \midrule
%    Dendrite & Input terminal  & $\sim$100     \\
%    Axon     & Output terminal & $\sim$10      \\
%    Soma     & Cell body       & up to $10^6$  \\
%    \bottomrule
%  \end{tabular}
%\end{table}

\bibliographystyle{unsrt}
\bibliography{neurips}

\newpage
\appendix
\section{Supplementary Material} \label{supp}
\subsection{Robot extension}
\begin{figure}[h!]
  \includegraphics[width=0.4\textwidth]{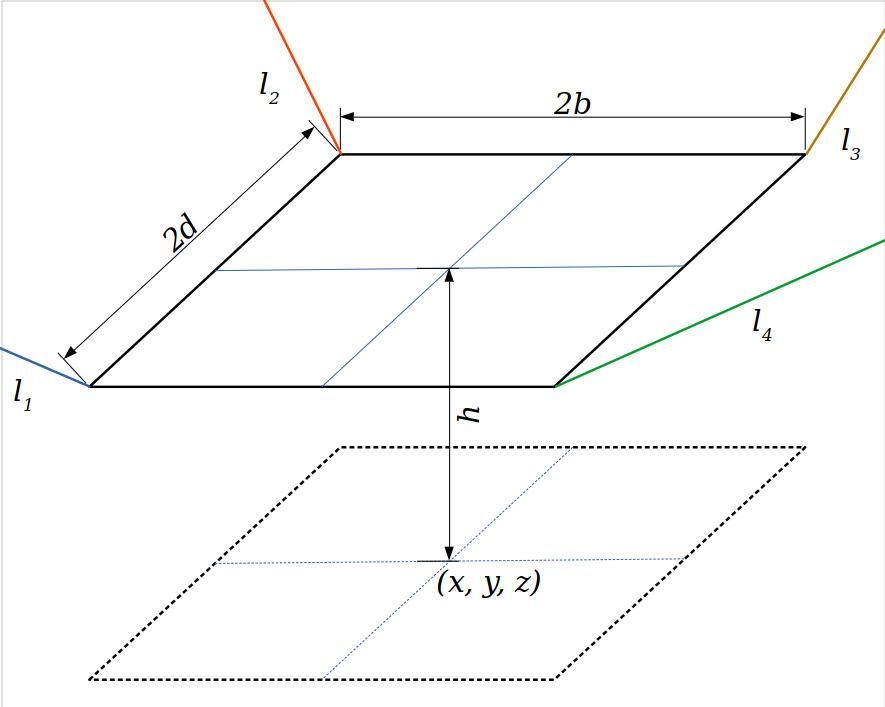}
  \centering
  \caption{Simulated environment setup.}
  \label{fig:robot-rect}
\end{figure}

The robot is considered as a box of dimensions 
$2b\times2d\times h$, with the strings being attached to the corners of the top face, see \cref{fig:robot-rect}. $(x,y,z)$ is the coordinate of the centroid of the bottom face, where the robot's claw (or other appendage of interest) is attached. Then, calculating the distance between the corners of the top face and the pulleys, and equating with the string lengths, we have:
\begin{align}
\begin{aligned}
L_1^2 &= (x - b)^2 + (y - d)^2 + (H - h - z)^2 \\
L_2^2 &= (x - b)^2 + (D - d - y)^2 + (H - h - z)^2 \\
L_3^2 &= (B - b - x)^2 + (D - d - y)^2 + (H - h - z)^2 \\
L_4^2 &= (B - b - x)^2 + (y - d)^2 + (H - h - z)^2
\end{aligned}\label{eqn:distances}
\end{align}

Subtracting pairs of the equations in \cref{eqn:distances}, we can 
get $x$ and $y$ in terms of the $L_i$:
\begin{align}
y = \frac{1}{2}\left(\frac{L_1^2 - L_2^2}{D - 2d} + D\right) &&
x = \frac{1}{2}\left(\frac{L_2^2 - L_3^2}{B - 2b} + B\right)
\end{align}
These values can be used in any of the equations in 
\cref{eqn:distances} to get $z$. Now, treating the robot as a point, 
$b, d$ and $h$ vanish, giving us:
\begin{align}
y = \frac{L_1^2 - L_2^2 + D^2}{2D} &&
x = \frac{L_2^2 - L_3^2 + B^2}{2B}
\label{eqn:xy}
\end{align}
Combining these with \cref{eqn:lengths}, we can obtain 
$(x, y, z)$ from the rotations $\Delta\theta_i$. \Cref{eqn:lengths,eqn:distances,eqn:xy} allow us to 
determine both the rotations needed to reach a particular point, as 
well as the location that will be reached by a set of rotations.
\newline

% Get to a new page here
%\newpage
\subsection{Data generation}
\begin{figure}[h!]
  \includegraphics[width=0.5\textwidth]{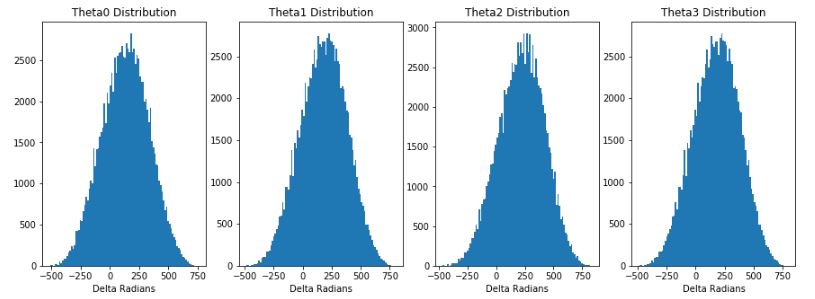}  
  \centering
  \caption{Delta Theta Distribution for data generated}
  \label{fig:delta}
\end{figure}

\subsection{Model evaluation}
\Cref{tbl:results} shows the $R^2$ score comparison. However, this is not the best way to judge if the model has generalized. In order to evaluate that, it is important that we check Error vs $Y(true)$ patterns and also the distribution of error. For comparison, we take GradBoost Model vs Neural Network Model.

\begin{table}[h!]
\centering
\scalebox{0.8}{
\begin{tabular}{ |p{6cm}|p{1cm}|p{1cm}|p{1cm}|p{1cm}|p{1cm}|} 
 \hline
 \multicolumn{6}{|c|}{Model performance metrics} \\
 \hline
 \textbf{Algorithm} & \multicolumn{4}{|c|}{\textbf{Test (r2 score)}}&\textbf{MCU}\\
 \hline
   & \textbf{Theta0} & \textbf{Theta1} & \textbf{Theta2} & \textbf{Theta3} &  \\
 \hline
 Linear Regression & 0.9245 & 0.9310 & 0.9393 & 0.9315 & No \\
 %\hline
 SGD ElasticNet & 0.9237 & 0.9300 & 0.9391 & 0.9307 & No \\
 %\hline
 DecisionTree Regressor & 0.9901 & 0.9921 & 0.9952 & 0.9922 & No \\
 %\hline
 Gradient Boosting & 0.9960 & 0.9939 & 0.9931 & 0.9932 & No \\
 %\hline
 XGBoost & 0.9874 & 0.9826 & 0.9827 & 0.9827 & No \\
 %\hline
 Neural Network (Fully Connected) GPU & \textbf{0.9985} & \textbf{0.9972} & \textbf{0.9990} & \textbf{0.9989} & Yes \\
 %\hline
 Neural Network (Fully Connected) MCU & 0.9742 & 0.9733 & 0.9761 & 0.9756 & Yes \\
 \hline
\end{tabular}}
\newline
\caption{Test scores and MCU compatibility}
\label{tbl:results}
\end{table}

\subsection{Error Distribution Inference}
The error distribution of GradBoost is much larger in comparison to that of neural network model. The Neural Network model provides a much sharper distribution of error around mean $0$ (zero). Lastly, there seems be a non-linear pattern between $Y (true value)$ and the error for GradBoost model which the algorithm is missing to map completely. However, for the NN model, the error is more inclined towards being purely random.

\begin{figure}[h!]
  \centering
  \includegraphics[width=0.5\textwidth]{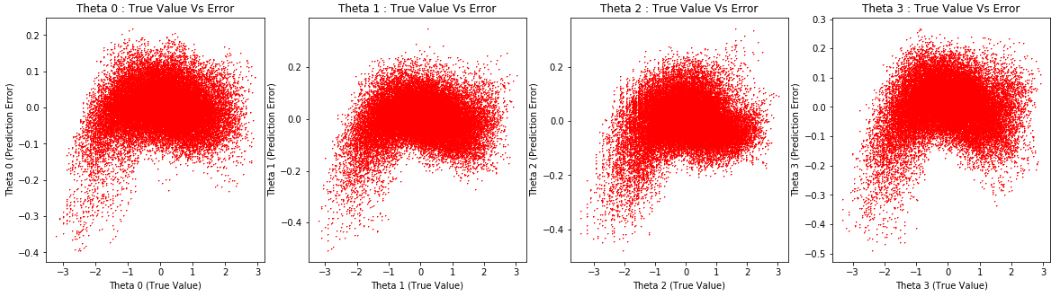}
  \includegraphics[width=0.5\textwidth]{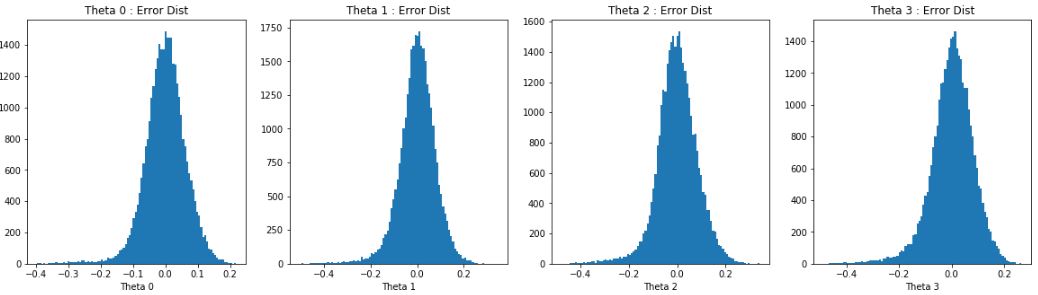}
  \label{gradboost}
  \caption{GradBoost Error}
\end{figure}

\begin{figure}[h!]
  \centering
  \includegraphics[width=0.5\textwidth]{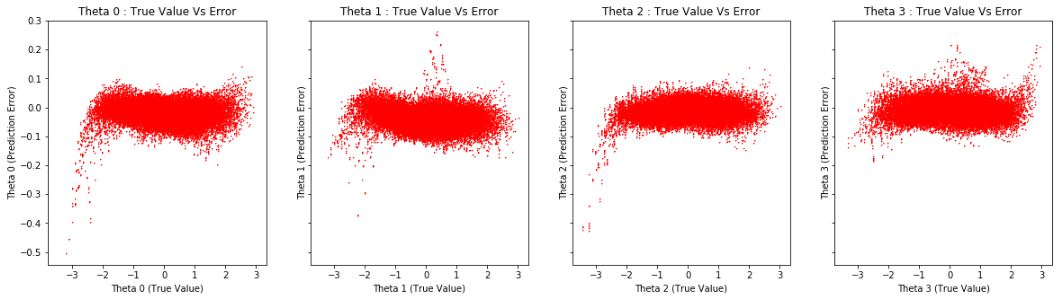}
  \includegraphics[width=0.5\textwidth]{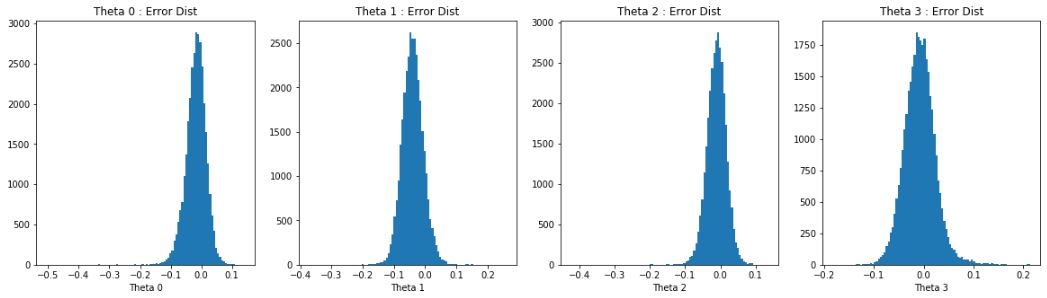}
  \label{nnmodel}
  \caption{Neural Network Error}
\end{figure}

\end{document}